\documentclass[sigconf,authorversion]{acmart}

\usepackage{balance}
\usepackage{listings}
\usepackage{placeins}
\usepackage{enumitem}
\usepackage{microtype}

\newcommand{\keyInstrument}{Computer Science Attitudes Scale}

\newlist{researchquestions}{enumerate}{1}
\setlist[researchquestions]{label*=\textbf{RQ\arabic*.}}

\newlist{openquestions}{enumerate}{1}
\setlist[openquestions]{label*=\textbf{PSQ\arabic*.}}

\AtBeginDocument{%
  \providecommand\BibTeX{{%
    \normalfont B\kern-0.5em{\scshape i\kern-0.25em b}\kern-0.8em\TeX}}}

\copyrightyear{2023}
\acmYear{2023}
\setcopyright{acmlicensed}\acmConference[ITiCSE 2023]{Proceedings of the 2023 Conference on Innovation and Technology in Computer Science Education V. 1}{July 8--12, 2023}{Turku, Finland}
\acmBooktitle{Proceedings of the 2023 Conference on Innovation and Technology in Computer Science Education V. 1 (ITiCSE 2023), July 8--12, 2023, Turku, Finland}
\acmPrice{15.00}
\acmDOI{10.1145/3587102.3588854}
\acmISBN{979-8-4007-0138-2/23/07}

\begin{document}

%\title{Using Sensor-Based Programming to Improve Interest in Computing for Students from Underrepresented Groups}

\title[Using Sensor-Based Programming to Improve Self-Efficacy and Outcome Expectancy]{Using Sensor-Based Programming to Improve Self-Efficacy and Outcome Expectancy for Students from Underrepresented Groups}

\author{Hussel Suriyaarachchi}
\orcid{0000-0002-8026-2523}
\affiliation{
\institution{Augmented Human Lab \\ National University of Singapore}
  \country{Singapore}
}
\email{hussel@ahlab.org}

\author{Alaeddin Nassani}
\orcid{0000-0002-3587-5523}
\affiliation{%
  \institution{Augmented Human Lab \\ University of Auckland}
  \city{Auckland}
  \country{New Zealand}
}
\email{alaeddin@ahlab.org}

\author{Paul Denny}
\orcid{0000-0002-5150-9806}
\affiliation{%
  \institution{School of Computer Science \\ University of Auckland}
  \city{Auckland}
  \country{New Zealand}
}
\email{paul@cs.auckland.ac.nz}

\author{Suranga Nanayakkara}
\orcid{0000-0001-7441-5493}
\affiliation{%
\institution{Augmented Human Lab \\National University of Singapore}
  \country{Singapore}
}
\email{suranga@ahlab.org}

\renewcommand{\shortauthors}{Suriyaarachchi et al.}


\begin{abstract}
Knowledge of programming and computing is becoming increasingly valuable in today's world, and thus it is crucial that students from all backgrounds have the opportunity to learn.  As the teaching of computing at high-school becomes more common, there is a growing need for approaches and tools that are effective and engaging for all students.  Especially for students from groups that are traditionally underrepresented at university level, positive experiences at high-school can be an important factor for their future academic choices.  In this paper we report on a hands-on programming workshop that we ran over multiple sessions for M\={a}ori and Pasifika high-school students who are underrepresented in computer science at the tertiary level in New Zealand.  
In the workshop, participants developed Scratch programs starting from a simple template we provided. 
In order to control the action in their programs, half of the participants used standard mouse and keyboard inputs, and the other half had access to plug-and-play sensors that provided real-time environmental data. 
We explore how students' perceptions of self-efficacy and outcome expectancy -- both key constructs driving academic career choices -- changed during the workshop and how these were impacted by the availability of the sensor toolkit. 
We found that participants enjoyed the workshop and reported improved self-efficacy with or without use of the toolkit, but outcome expectancy improved only for students who used the sensor toolkit. 

\end{abstract}




\maketitle

%- - - - - - - - - - - - - - - - - - - - - - - - - -
%                 INTRODUCTION
%- - - - - - - - - - - - - - - - - - - - - - - - - -
\section{Introduction}
As knowledge of computing and programming become increasingly important in the modern world, it is essential that students from all backgrounds have the opportunity to learn and excel in this field.  In recent years, many countries have made efforts to develop and adopt national curricula in recognition of the importance of building a strong foundation in this area for the next generation of students \cite{duncan2015pilot, falkner2019international, brown2014restart}.%, seehorn2011csta}.  
However increasing engagement and success within the discipline for students from traditionally underrepresented groups remains a persistent global problem \cite{lunn2021exploration, newhall2014support, mcbroom2020understanding}.  

In the New Zealand context, M\={a}ori and Pasifika students are underrepresented at the tertiary level in computer science and related fields~\cite{McAllister2022}. One potential barrier to the participation and success of M\={a}ori and Pasifika students in tertiary-level computing is the lack of culturally relevant learning materials~\cite{gaddam2018culturally}.  
However, the development of such materials can be difficult for educators \cite{yuen2016culturally}, and directly affects existing students. %only. 
The core problem of attracting underrepresented students into tertiary-level computing programmes from high-school remains.  
Expectancy-value theory suggests that an individual's academic career choices are influenced by their \emph{expectations of academic success} and the \emph{value they perceive in academic tasks} in the subject area~\cite{Rachmatullah2020}.  These two attitudinal constructs -- self-efficacy and outcome expectancy -- therefore play a key role in shaping the goals that students set for themselves and the choices they make.  Activities that improve a student's expectations of academic success, as well as the inherent value the student associates with this success, may be valuable for attracting students to further study.

In this paper, we explore the impact of a hands-on programming workshop on the self-efficacy and outcome expectations of M\={a}ori and Pasifika high-school students in New Zealand.  During the workshop, participants developed programs in Scratch using different input modalities.  In particular, we conduct a controlled study to determine whether the use of a sensor toolkit that provides real-time environmental data as program input has an effect on either of the attitudinal dimensions we measure.  Our hypothesis is that empowering students by providing additional options for controlling program behaviour, such as through data from the environment, will increase the value that students perceive in being successful with the task.  We measure self-efficacy and outcome expectancy with respect to computer science and programming using the recently validated \keyInstrument~\cite{Rachmatullah2020}.  Our evaluation of the workshop is guided by the following research questions:

\begin{researchquestions}
    \item How do measures of self-efficacy and outcome expectancy change after students participate in a hands-on programming workshop, and does the use of a sensor-based toolkit impact these changes?
    \smallskip
    \item What do students report as being the most enjoyable aspect of the workshop activity, and how does the sensor toolkit impact decisions about controlling program behaviour?
    \smallskip
\end{researchquestions}

Our findings suggest that the workshop was successful, improving feelings of self-efficacy for participants regardless of whether the sensor toolkit was used.  
However outcome expectancy improved only for students who used the sensor toolkit. We discuss the implications of this work for improving interest in studying computer science for traditionally underrepresented cohorts.

%- - - - - - - - - - - - - - - - - - - - - - - - - -
%                 RELATED WORK
%- - - - - - - - - - - - - - - - - - - - - - - - - -
\section{Related Work}

\subsection{Self-efficacy and outcome expectancy}

Psychologists define self-efficacy as the belief in one's ability to succeed~\cite{Bandura1978}. Such belief with respect to academic tasks has been linked to better performance, engagement and career aspirations \cite{Frank1988}. Thus, strategies for improving student self-efficacy are of great interest to educators~\cite{loksa2022metacognition}.  One promising direction appears to be that creating technology through `making' activities (e.g. programming, electronics and 3D printing) can be effective in boosting self-efficacy in science and increasing interest in STEM fields~\cite{Schlegel2019}.

Outcome expectancy, on the other hand, refers to an individual's belief about the likelihood that a specific outcome will occur as a result of their actions. This is closely linked to self-efficacy, as the belief in one's ability to perform a task can influence how much effort or time they are willing to invest in order to achieve the outcome.  Outcome expectancy has also been found to be an important predictor of underrepresented groups' engagement in STEM~\cite{Lent1994, Hanson2004}. Indeed, studies have shown that students from underrepresented groups, such as women and minorities in STEM subjects~\cite{Fouad2017, Fong2021}, often have lower outcome expectancies for their performance.  This may contribute to poor representation within STEM fields, and provides strong motivation for work that aims to improve outcome expectancy beliefs.

Prior work has used a variety of instruments for measuring self-efficacy and outcome expectancy for computer science (CS) students.% under various scale names. 
The CS Attitude Survey~\cite{Wiebe2003} is an early instrument related to non-cognitive factors, however it has been criticised for its lack of construct validation and outdated psychological models. Korkmaz et al. proposed the Computational Thinking Scale ~\cite{Korkmaz2017} as an alternative 29-item instrument.  This has been validated with undergraduate students, measuring factors such as algorithmic thinking and creativity, however it does not have a theoretical foundation involving non-cognitive constructs such as self-efficacy or outcome expectancy.
Kukul et al.~\cite{Gokcearslan2017} and Tsai et al.~\cite{Tsai2019} have both developed Computer Programming Self-Efficacy Scale instruments 
%consisting of multiple items 
targeting 
%self-efficacy related to 
different facets of computational thinking (CT). Both instruments take a broad approach to structure item sets around CT/CS constructs, but they are long instruments which can create reliability and validity problems due to logistical time pressures or respondent fatigue.

The most recent instrument, the \keyInstrument~developed by Rachmatullah et al.~\cite{Rachmatullah2020} is a concise, reliable and validated instrument for measuring self-efficacy and outcome expectancy in educational research. It is designed specifically for computer science, and addresses some shortcomings of other attitudinal instruments.
A contribution of the current work is we present one of the first uses of the 
Computer Science Attitudes Scale
to measure self-efficacy and outcome expectancy in 
high school students from traditionally underrepresented groups. 

\subsection{Underrepresented groups in computing}

The underrepresentation of certain groups in STEM disciplines has been a long-standing issue at high-school and tertiary level, especially in computing~\cite{Guzdial2012, margolis2012beyond}. Studies have found that women, ethnic minorities, and individuals with disabilities are significantly underrepresented in STEM fields. Factors that contribute to low representation include gender and racial stereotypes, lack of role models, and limited access to educational resources. 

Ethnic minorities, such as M\={a}ori and Pasifika in the New Zealand context, are also significantly underrepresented in STEM fields~\cite{McAllister2022}. Studies  have found that ethnic minorities are less likely to pursue STEM degrees and careers, and are more likely to drop out of STEM programs. 
Mentorship can foster scientific identity and career pathways for underrepresented minorities in STEM fields. Atkins et al~\cite{Atkins2020} showed that research mentorship was associated with a stronger scientific identity and suggested that tailored mentoring approaches and research-focused mentoring are needed to support the development of future STEM leadership.

\subsection{Sensor-enabled block programming}

The introduction of block-based programming environments has helped to make programming more accessible to a wide range of students. %made it more accessible to students. 
Most tools focus on writing programs to control a device and observe code execution~\cite{klassner2003lego, sentance2017creating, mellis2007arduino}. 
Block-based programming tools have been used to 
teach cryptography algorithms in the classroom~\cite{Percival2022}
as well as 3D printing~\cite{Schlegel2019} 
and educational robot programming~\cite{ARSLAN2020,Coban2020}.
Weintrop et al~\cite{Weintrop2018} directly compared block-based programming to text-based programming, and observed that female students and students from historically underrepresented minorities benefited the most from the block-based format. This suggests that the choice of programming environment and modality can positively affect those who are from traditionally marginalised groups.

Compared to directly controlling an external device, relatively little attention has been paid to using real-time sensor hardware for input to control general program behaviour. 
Previous research that has explored the use of environmental sensors in block-based programming environments has shown that they can be highly motivating for young students \cite{Suriyaarachchi2022}. 

In the current work, we explore the use of such an environment for 14-15 year old students, from groups who are traditionally underrepresented in computing at the tertiary level, who are close to making decisions about their future studies and careers.

%- - - - - - - - - - - - - - - - - - - - - - - - - -
%                 METHODOLOGY
%- - - - - - - - - - - - - - - - - - - - - - - - - -
\section{Methodology}

\subsection{Programming environment \& sensor toolkit}
We used the Scratch\footnote{\url{https://scratch.mit.edu/}} programming environment in our study, given its suitability across a wide age range and the fact that we had little information ahead of time regarding the prior programming experience of our participants. 
Scratch provides a convenient environment in which students can produce an interactive program quickly and with little need for support as well as access to a rich collection of programming blocks and the ability for users to control actions within a program using the keyboard and mouse. 

\begin{figure}[t]
    \centering
    \includegraphics[width=\linewidth]{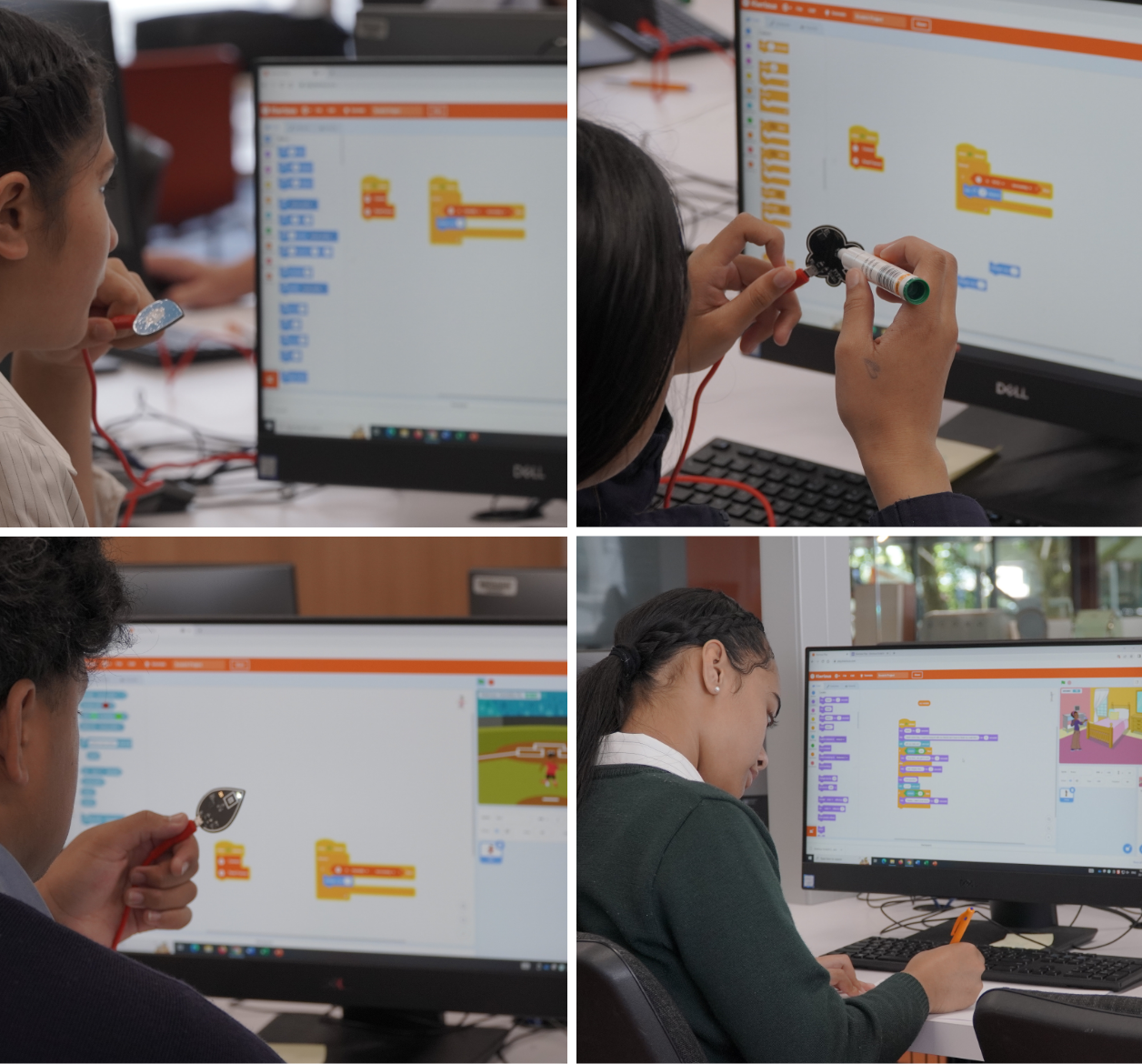}
    \vspace{-15pt}
    \caption{Students engaging in our hands-on programming workshop where they used a sensor toolkit that provided real-time environmental data as input to their Scratch programs.}
    \label{fig:session}
    \vspace{-10pt}
\end{figure}

We used a derivative of the Scratch interface implemented by Suriyaarachchi et al.~\cite{Suriyaarachchi2022, suriyaarachchi2022primary}, containing an extension that enables the use of environmental data from their custom plug-and-play sensor toolkit when programming.
Figure \ref{fig:session} shows students actively engaging with the sensor toolkit during our workshop. The toolkit comprises sensors such as humidity, surface temperature and visible light, which could be used as real-time input to a running program.
Connectivity to Scratch is supported through the computer's USB port, making for a seamless plug-and-play experience that avoids the need for additional software or setting-up, which would likely interfere with the study as sessions were scheduled back to back. The sensors introduce a new modality of input, allowing students to create interactive programs that can respond to changes in their environment.

The ability to utilise all parts of the standard Scratch platform, independent of sensor availability, was an important aspect of our study design.  This enabled us to focus on measuring the effect of the sensors as the only variable between the control and experimental groups. We explain this design in the following sections. 

\subsection{Participants}

\begin{figure*}
    \vspace{-10pt}
    \centering
    \includegraphics[width=\textwidth]{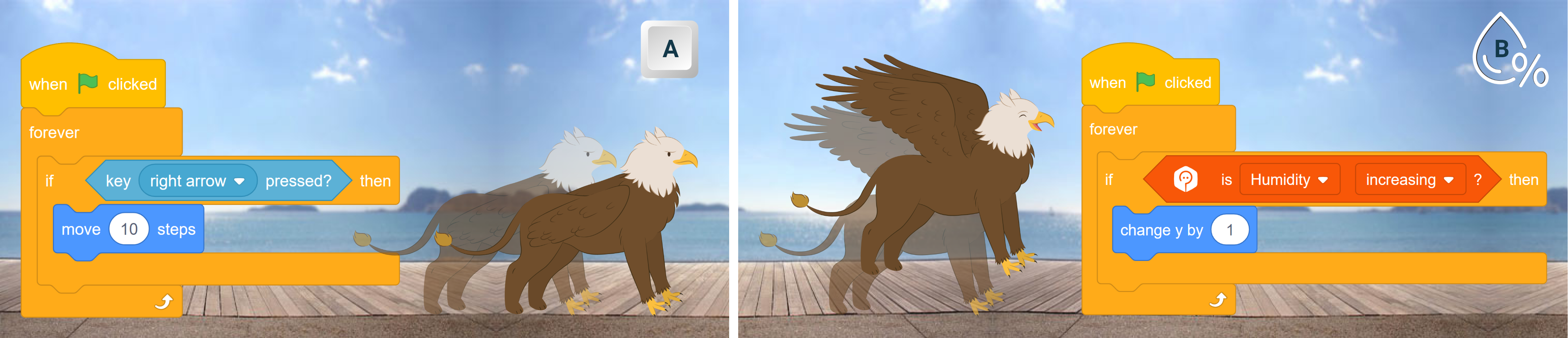}
    \vspace{-15pt}
    \caption{A) Sample Scratch program demonstrated to the control (non-sensor) group. The sprite is controlled by keyboard input and moves to the right when the right arrow key is pressed; B) Sample program provided to the experimental (sensor) group relies on input from the sensor toolkit to control the sprite instead. An increase in the humidity will move the sprite upward.}
    \label{fig:demo}
    \vspace{-5pt}
\end{figure*}

Our workshop and study were conducted as part of an outreach campaign called \textit{Future Me}, organised centrally by The University of Auckland, New Zealand.
The aim of the campaign was to encourage Year 10 (typical ages 14–15) M\={a}ori and Pasifika students, in the last two or three years of high-school study, to continue their education at the tertiary level. We hosted four groups of high school students ($N=49$) during the campaign. 
Fewer than 60\% of participants used a computer on every or most days, and around 20\% reported using a computer not very often or never. These patterns are consistent with prior work that has explored technology use in underrepresented communities~\cite{google2016diversity}.

\subsection{Study Procedure}
The workshop consisted of four 40-minute sessions, one for each group of students.
%, over which our study was performed.  
Each 40-minute session composed of a short introduction to programming with Scratch (10 minutes), an open-ended coding activity (20 minutes), and data collection (10 minutes). We randomly selected two of the groups to serve as the experimental condition, in which the sensor toolkit was distributed at the beginning of the workshop and collected at the end. The 24 students, in two groups, served as the experimental condition and were shown how to connect the sensor to the USB port and how to add a sensor-specific block to their program, and thus were able to use sensor-based data for controlling the action of sprites in their programs.  The remaining 25 students, in the other two groups, served as the control condition and were free to create programs in which the sprites could be controlled with the keyboard and mouse. The Scratch extension we use in this work provides full access to standard Scratch blocks when the sensors are not connected. Thus, students in both groups used an identical interface in all respects, with the exception of the sensors and sensor-specific blocks available only to participants in the experimental group.

\subsubsection{Introduction to Scratch}

The introduction provided all participants with a basic overview of connecting blocks in Scratch, and aimed to familiarise them with navigating the interface. 
To facilitate this, a short sample program was provided to students which illustrated the core components and operations of Scratch.
Figure~\ref{fig:demo} shows the program that was demonstrated, along with the blocks, which cover concepts such as Sprite control, event handling and logical operations.  The only difference between the sample program that was used for students in the control and experimental conditions was in how the Sprite was controlled by the user.  
The input source for the control group's program, signified by a conventional keypress (Figure~\ref{fig:demo}A), is substituted by the sensor block for the experimental group (Figure~\ref{fig:demo}B) that awaits a rise in the measured humidity level to change the sprite's position on the screen. The behaviour of the Sprite in both conditions is comparable, varying only in the input modality. 

\subsubsection{Programming Activity}
We proceeded to occupy the students in a 20-minute activity where they could explore programming with the various input options available to their respective groups. While students had complete autonomy in what they chose to create, their programs were required to contain at least one aspect controlled by an input representative of their study group. Moderating the students' use of input mechanisms, namely, the traditional keyboard/mouse in the control group and sensors in the experimental group, was necessary to generate a valid comparison for our evaluation. Five instructors who were experienced in Scratch were present during the activity to assist students and verify that their programs  satisfied our requirements. We suggested that students ``tinker''~\cite{dong2019defining} with the sample program should they find it challenging to create a new program on their own.

\subsubsection{Data Collection}

We used a pre- and post-questionnaire containing Likert-scale questions to evaluate the sensor-based toolkit used during the programming workshop. It consisted of ten questions (see Table~\ref{tbl:questions}): three on Self-Efficacy (I1-I3), six on Outcome Expectancy (I4-I9) and one on confidence in programming with Scratch (Q1). The items I1-I9 were adopted from the \keyInstrument~\cite{Rachmatullah2020}.
The pre-questionnaire was administered before the session's introduction, and the post-questionnaire, soon after students completed their programming activity.
Responses to all items were recorded on a 5-point Likert scale (where 1 = strongly disagree and 5 = strongly agree).
Additionally, two open-ended questions appeared in the post-study questionnaire: 
\begin{openquestions}
    \smallskip
    \item Tell us what you enjoyed most about today's activity.
    \smallskip
    \item Tell us a little about your Scratch project. How did you control your character(s)? Why did you program them that way?
\end{openquestions}
%These questions gather more detailed descriptions of the students' experiences during the workshop. The s
Students were encouraged to attempt all questions to the best of their ability based on their sentiments at the time and were reassured that there were no right or wrong answers.

\subsubsection{Data Analysis}
\label{sec:methods:analysis}

We received completed pre/post questionnaires from 45 (out of 49) students who participated in our study. For each student, we inferred their perceived self-efficacy and outcome expectancy by calculating the average of their responses to the questionnaire items I1-I3 and I4-I9, respectively. This was done for both the pre- and post-questionnaire, resulting in paired measurements of self-efficacy and outcome expectancy for each student. We divided the collected data into two groups: control (non-sensor; $n=25$) and experimental (sensor; $n=20$). A one-sample paired t-test on the differences in the paired data within each group was used to determine if these changes were significant. Despite the ordinal nature of Likert scale responses, prior work by Vieira et al.~\cite{vieira2016t} found t-tests to be an appropriate statistical method for comparing such data. To further validate the use of a t-test in our analysis, we performed a Shapiro-Wilk test, which confirmed that our data was normally distributed.

We conducted a qualitative analysis to get insight into what students enjoyed about the session. 
Following the guidelines for thematic analysis outlined by Braun and Clarke~\cite{braun2006using}, we tagged responses to the open-ended question \textit{PSQ1} and synthesised these tags into high-level themes.
When presenting these findings in Section \ref{sec:result:open}, we report on some of the free-form responses to question \textit{PSQ2} and illustrate how these themes were reflected in the students' programming tasks.

\begin{table*}[t]
    \vspace{-5pt}
    \centering
    \caption{Average scores (5-point Likert) of Self-Efficacy and Outcome Expectancy before (pre) and after (post) the task for each sensor group. $\mu_\textit{diff}=\textit{post-pre}$.}
    \vspace{-10pt}
    \includegraphics[width=\textwidth]{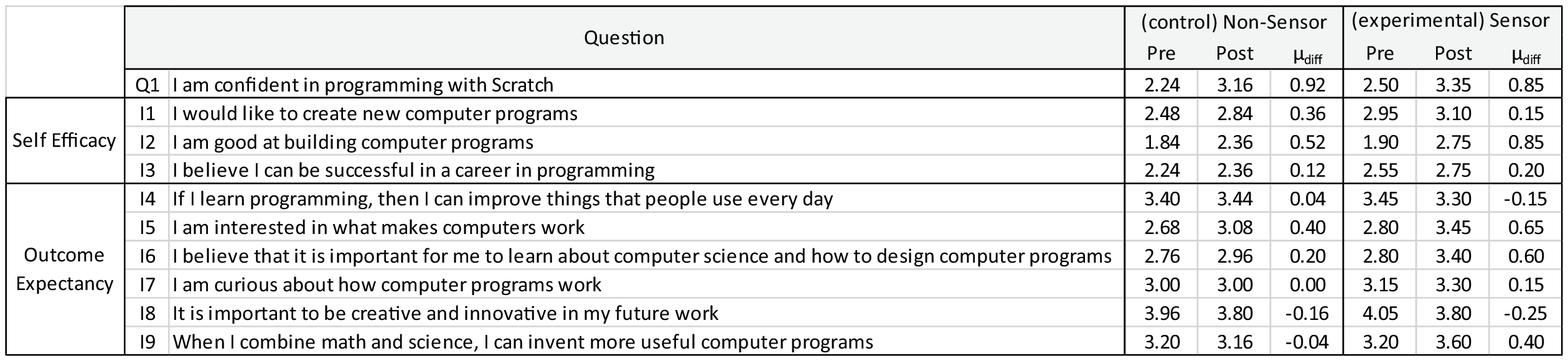}
    \label{tbl:questions}
    \vspace{-15pt}
\end{table*}

%- - - - - - - - - - - - - - - - - - - - - - - - - -
%                 RESULTS & DISCUSSION
%- - - - - - - - - - - - - - - - - - - - - - - - - -
\section{Results \& Discussion}
\label{sec:result}

In Section ~\ref{sec:result:subjective} we present the results from the pre- and post- questionnaires in order to answer RQ1. In Section~\ref{sec:result:open} we present findings from the open-ended questions in order to address RQ2. Finally, we describe limitations and suggest avenues for future work in Section~\ref{sec:result:limitation}.

\subsection{Student Attitudes and Perceptions}
\label{sec:result:subjective}

Table \ref{tbl:questions} lists the wording of each question as it appeared on the questionnaires, and presents the average response for each question calculated across both groups (non-sensor and sensor). 
We observe an increase in $\mu_\textit{diff}$ for Q1 concerning the confidence to program with Scratch in both groups. The non-sensor group had a higher $\mu_\textit{diff}$ (0.92) than the sensor group; however, the sensor group had higher \textit{pre} (2.5) and \textit{post} (3.35) values.  In general, most students were not initially very confident with Scratch (the pre-values represent an average response somewhere between ``Disagree'' and ``Neutral'').  As would be expected, students generally felt more confident programming in Scratch after completing the workshop, regardless of which input modality they used.  
Similarly for I1-I9, we see a general increase in the overall average of $\mu_\textit{diff}$ for the sensor group (0.29), which was greater than the non-sensor group (0.16). 
One possible explanation for this increase could be that the sensor-based toolkit provided a hands-on learning experience, making programming concepts easier to understand, leading to an increase in their self-efficacy and outcome expectancy.

\begin{table}[b]
    \vspace{-10pt}
    \caption{Analysis of the pre and post results within each group using one-sample paired t-test. $\mu_\textit{diff}=\textit{post-pre}$, *= significant with $p<0.05$.}
    \vspace{-10pt}
    \centering
    \includegraphics[width=\linewidth]{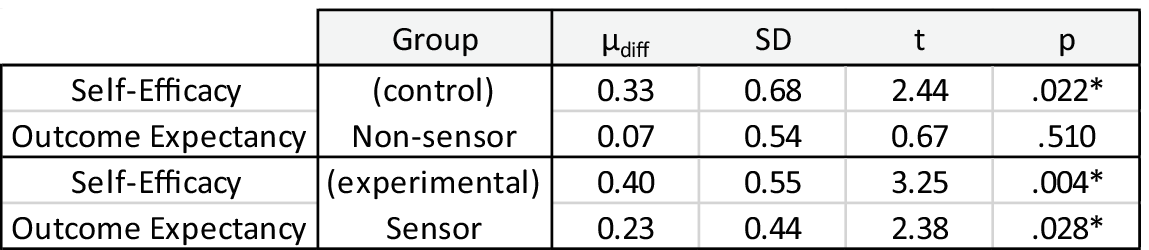}
    \label{tbl:stat}
\end{table}

The results of our statistical analysis (see Table~\ref{tbl:stat}) explore the changes in self-efficacy and outcome expectancy among students of the control (non-sensor) and experimental (sensor) groups.
For the control group, we found that self-efficacy improved significantly between pre and post scores $t(24)=2.44, p=0.022$. However there was no significant difference in outcome expectancy between pre and post scores $t(24)=0.67, p=.510$. 
For the experimental group, we found that both self-efficacy ($t(19)=3.25, p=.004$) and outcome expectancy ($t(19)=2.38, p=.028$)  improved significantly between pre and post scores.  

There can be a difference between self-efficacy and outcome expectancy when a participant believes they are capable of performing a task but they have a low expectation of the outcome they will receive. This can happen if the participant perceives the outcome to be out of their control or influenced by external factors. 
In our study, participants using  the sensors were able to interact with the environment around them, and this may have improved their perceptions of the value of the programming activity.

\subsection{Open-response feedback}
\label{sec:result:open}

We observed a wide variety of responses to \textit{PSQ1}, which asked students to describe what they enjoyed most about the workshop.  Some of these were short generic statements, including ``It was fun and interesting'', which we coded with the tag `fun'. From the tags assigned to the more substantial comments, three central themes emerged that were common across responses from students in both groups. We use the terms \emph{learn}, \emph{create} and \emph{interact} to describe these three themes.  Although the three themes were common to both groups, students tended to view the opportunities to learn, create and interact in different ways, which we expand on below.

\subsubsection{Opportunities for Learning}
Students in the control group viewed learning as acquiring knowledge about programming and using Scratch. Exploring the ``\emph{different things you can do on Scratch}'' and ``\emph{figuring out how to code}'' were particularly memorable experiences for novice users. 
Some students with prior experience using Scratch indicated they enjoyed the opportunity to develop their skills further such as  
``\emph{being able to extend my ability in computer programming and trying new codes out}''.  
Requiring students to produce a functional program that incorporated standard input as a control mechanism may have motivated a deeper exploration of Scratch's native functionalities. As a result, they may have found the challenge of working within these constraints to be engaging and rewarding.

In contrast, students in the experimental group tended to explicitly refer to the sensors when expressing what they had enjoyed learning.
With sensors being a novelty to students, there was much interest in getting to know ``\emph{how the sensors worked}'' and developing an ``\emph{understanding [of] how a program is connected through the sensors}''. Students also noted the unique capabilities offered by environmental data and the potential of the technology they were learning to use, ``\emph{I was able to do something different that I don't always do in school}''. 

While both groups valued the opportunity to learn programming and Scratch, the intuitive learning contexts provided by the sensors were appealing to the experimental group.
Thus, the popularity of sensors in the learning experiences reported by students highlight the importance of adopting hands-on and real-world approaches.

\subsubsection{Opportunities for Creating}
Students in both groups appeared to enjoy the creative aspects of the workshop. For those in the control group, this enjoyment arose from the creative freedom supported by the workshop. Students voiced enthusiasm for the ability to translate their imagination into a working program, ``\emph{[I enjoyed] having a go at creating my own character and story on Scratch. It was fun!}''. The open-ended structure of our programming activity was a key aspect that promoted these creative processes and the positive remarks that followed. References to the lack of constraints and the freedom to explore and experiment with different ideas were observed in comments such as ``\emph{it gave me a chance to do what I wanted to do and create}''.

Students in the experimental group enjoyed creating by drawing personal and cultural associations to the sensor data they worked with and the environmental conditions they were measuring. 
The ties to various elements of nature established by the sensors inspired students to create programs that reflected their personal connection to them. For example, in their response to \textit{PSQ2}, a student who used the humidity sensor and discovered its relationship to water, was reminded of the ocean, ``\emph{I enjoy the sea. It is the place I go to feel calm and at peace. I did my program based on the ocean because it is my safe haven}''.
Furthermore, the evocative shapes of the sensors were also reflected on by students. In one instance, a student used the native term ``moana'' (water) to describe the humidity sensor (which was designed to resemble a water droplet).

\balance

\subsubsection{Opportunities for Interacting}
Students in the control group reported interactions with the instructors as the most enjoyable experience of the workshop. With respect to the available input modalities, the ``\emph{limited language of Scratch}'' may have prompted this, as students had to frequently rely on instructors' assistance to navigate the discrete controls of the keyboard and mouse. Representative comments include: ``\emph{What I enjoyed the most about today's activity is interacting with the crew}'' and ``\emph{I liked how nice the people [instructors] are}''. Indeed, as recent literature suggests~\cite{denny2019research}, the role of instructors in shaping a student's experiences is  just as important as the interactive elements available during programming.

In the experimental group, students enjoyed manipulating their immediate physical environment as a way of interacting with the programs they created.
Themes revealing playful interactions with the sensors appeared in twelve of the twenty responses to \textit{PSQ1}.
Actions such as ``breathing'' and ``blowing'' on sensors symbolised the impact of a tangible entity in fostering an authentic interplay between the physical and virtual worlds.
As students continued to enjoy ``\emph{the different ways I can control actions of a character [using sensors]}'' and ``\emph{observe their results in real-time}'', their accounts of how the programs were implemented (\textit{PSQ2}) followed suit. One student writes: ``\emph{[My program] is a balloon in which you use a VOC [Volatile Organic Compound] sensor, and when it increases, the balloon will rotate anti-clockwise, slide to a random position for 1 sec, then change colour}''.

\subsection{Limitations and Future Work}
\label{sec:result:limitation}

The goal of our study was to explore if the use of a sensor-based toolkit for measuring environmental data would meaningfully impact students' feelings of self-efficacy and outcome expectancy, both of which are known to be key drivers behind academic career choices.  
Our study ran in the context of
%We conducted our study in the context of 
a workshop designed for students who are typically underrepresented at tertiary level, given the growing importance of attracting students from these groups.

Although we observed significant increases in self-efficacy for both groups, and outcome expectancy for the experimental group, we are cautious in over-stating these findings for several reasons. 
The sample size ($n=45$) was relatively small and may not be representative of the larger population of underrepresented students. For example, we were unaware of how students were recruited and selected to participate in the workshops organised by the outreach campaign.  It is possible that a self-selection bias is present, where students who are particularly motivated may have been more likely to volunteer to take part. 
The research did not investigate other factors that may potentially influence self-efficacy and outcome expectancy such as access to resources, mentorship, or support from peers or family. Exploration of such factors would be a fruitful avenue for future work.  In addition, the study was limited to a single type of block-based programming environment (i.e. Scratch) and a single type of sensor toolkit and may not apply to other programming languages or sensor technology.

The instrument used to measure self-efficacy and outcome expectancy was entirely dependent on self-reported subjective responses.
Future work should focus on refining the methods (potentially adding objective measures such as completion time~\cite{Kesselbacher2019} or physiological data~\cite{Martin2021}) used to confirm the results and to measure self-efficacy and outcome expectancy. Specifically, a more rigorous random assignment process (we randomly selected two of the four groups to serve in each condition of our study, but we did not have insight into how students were assigned to these groups) and complementary measures to ensure reliability of  findings. 
Finally, our study was conducted over a short timeframe, as we were constrained by the length of the sessions that were scheduled as part of the outreach campaign. A longitudinal study should be conducted to measure the long-term impact of these interventions on enrolments into tertiary level study, although this can be challenging.

\section{Conclusion}

In this paper, we report on a hands-on programming workshop for high school students from M\={a}ori and Pasifika backgrounds.  We investigate how feelings of self-efficacy and outcome expectancy change during the workshop, and explore whether these are impacted differently based on the input modalities students use.  We found that the workshop was successful in improving students' perceptions of self-efficacy, with an increase in outcome expectancy observed specifically for those who used a sensor-toolkit for accessing environmental data. Additionally, students reported that the most enjoyable aspects of the workshop were learning, creating, and interacting. Overall, the use of sensor-based programming holds great potential in promoting inclusivity and equity in computer science education for underrepresented groups, particularly in terms of shaping students' academic choices through an improvement in self-efficacy and outcome expectancy.

\section*{Acknowledgements}
This work was supported by the Assistive Augmentation research
grant under the Entrepreneurial Universities (EU) initiative of New Zealand and the Tier 1 Ministry of Education (MOE) Academic Research Fund (AcRF).

\bibliographystyle{ACM-Reference-Format}
\bibliography{sample-base}


\begin{thebibliography}{41}

%%% the Web address.

\ifx \showCODEN    \undefined \def \showCODEN     #1{\unskip}     \fi
\ifx \showDOI      \undefined \def \showDOI       #1{#1}\fi
\ifx \showISBNx    \undefined \def \showISBNx     #1{\unskip}     \fi
\ifx \showISBNxiii \undefined \def \showISBNxiii  #1{\unskip}     \fi
\ifx \showISSN     \undefined \def \showISSN      #1{\unskip}     \fi
\ifx \showLCCN     \undefined \def \showLCCN      #1{\unskip}     \fi
\ifx \shownote     \undefined \def \shownote      #1{#1}          \fi
\ifx \showarticletitle \undefined \def \showarticletitle #1{#1}   \fi
\ifx \showURL      \undefined \def \showURL       {\relax}        \fi
% invisible to TeX
\providecommand\bibfield[2]{#2}
\providecommand\bibinfo[2]{#2}
\providecommand\natexlab[1]{#1}
\providecommand\showeprint[2][]{arXiv:#2}

\bibitem[ARSLAN and İŞBULAN(2020)]%
        {ARSLAN2020}
\bibfield{author}{\bibinfo{person}{Esra ARSLAN} {and} \bibinfo{person}{Onur
  İŞBULAN}.} \bibinfo{year}{2020}\natexlab{}.
\newblock \showarticletitle{{The Effect of Individual and Group Learning on
  Block-Based Programming Self-Efficacy and Robotic Programming Attitudes of
  Secondary School Students}}.
\newblock \bibinfo{journal}{\emph{Malaysian Online Journal of Educational
  Technology}} \bibinfo{volume}{9}, \bibinfo{number}{1} (\bibinfo{year}{2020}),
  \bibinfo{pages}{108--121}.
\newblock
\urldef\tempurl%
\url{https://doi.org/10.17220/mojet.2021.9.1.249}
\showDOI{\tempurl}


\bibitem[Atkins et~al\mbox{.}(2020)]%
        {Atkins2020}
\bibfield{author}{\bibinfo{person}{Kaitlyn Atkins}, \bibinfo{person}{Bryan~M.
  Dougan}, \bibinfo{person}{Michelle~S. Dromgold-Sermen},
  \bibinfo{person}{Hannah Potter}, \bibinfo{person}{Viji Sathy}, {and}
  \bibinfo{person}{A.~T. Panter}.} \bibinfo{year}{2020}\natexlab{}.
\newblock \showarticletitle{{“Looking at Myself in the Future”: how
  mentoring shapes scientific identity for STEM students from underrepresented
  groups}}.
\newblock \bibinfo{journal}{\emph{Int. Journal of STEM Education}}
  \bibinfo{volume}{7}, \bibinfo{number}{1} (\bibinfo{year}{2020}),
  \bibinfo{pages}{1--15}.
\newblock
\showISSN{21967822}
\urldef\tempurl%
\url{https://doi.org/10.1186/s40594-020-00242-3}
\showDOI{\tempurl}


\bibitem[Bandura(1978)]%
        {Bandura1978}
\bibfield{author}{\bibinfo{person}{Albert Bandura}.}
  \bibinfo{year}{1978}\natexlab{}.
\newblock \showarticletitle{{Self-efficacy: Toward a unifying theory of
  behavioral change}}.
\newblock \bibinfo{journal}{\emph{Advances in Behaviour Research and Therapy}}
  \bibinfo{volume}{1}, \bibinfo{number}{4} (\bibinfo{year}{1978}),
  \bibinfo{pages}{139--161}.
\newblock
\showISSN{01466402}
\urldef\tempurl%
\url{https://doi.org/10.1016/0146-6402(78)90002-4}
\showDOI{\tempurl}


\bibitem[Braun and Clarke(2006)]%
        {braun2006using}
\bibfield{author}{\bibinfo{person}{Virginia Braun} {and}
  \bibinfo{person}{Victoria Clarke}.} \bibinfo{year}{2006}\natexlab{}.
\newblock \showarticletitle{Using thematic analysis in psychology}.
\newblock \bibinfo{journal}{\emph{Qualitative Research in Psychology}}
  \bibinfo{volume}{3}, \bibinfo{number}{2} (\bibinfo{year}{2006}),
  \bibinfo{pages}{77--101}.
\newblock
\urldef\tempurl%
\url{https://doi.org/10.1191/1478088706qp063oa}
\showDOI{\tempurl}


\bibitem[Brown et~al\mbox{.}(2014)]%
        {brown2014restart}
\bibfield{author}{\bibinfo{person}{Neil C.~C. Brown}, \bibinfo{person}{Sue
  Sentance}, \bibinfo{person}{Tom Crick}, {and} \bibinfo{person}{Simon
  Humphreys}.} \bibinfo{year}{2014}\natexlab{}.
\newblock \showarticletitle{Restart: The Resurgence of Computer Science in UK
  Schools}.
\newblock \bibinfo{journal}{\emph{ACM Trans. Comput. Educ.}}
  \bibinfo{volume}{14}, \bibinfo{number}{2}, Article \bibinfo{articleno}{9}
  (\bibinfo{date}{June} \bibinfo{year}{2014}), \bibinfo{numpages}{22}~pages.
\newblock
\urldef\tempurl%
\url{https://doi.org/10.1145/2602484}
\showDOI{\tempurl}


\bibitem[{\c{C}}oban et~al\mbox{.}(2020)]%
        {Coban2020}
\bibfield{author}{\bibinfo{person}{Emre {\c{C}}oban},
  \bibinfo{person}{{\"{O}}zgen Korkmaz}, \bibinfo{person}{Recep {\c{C}}akır},
  {and} \bibinfo{person}{Feray {Uğur Erdoğmuş}}.}
  \bibinfo{year}{2020}\natexlab{}.
\newblock \showarticletitle{{Attitudes of IT teacher candidates towards
  computer programming and their self-efficacy and opinions regarding to
  block-based programming}}.
\newblock \bibinfo{journal}{\emph{Education and Information Technologies}}
  \bibinfo{volume}{25}, \bibinfo{number}{5} (\bibinfo{year}{2020}),
  \bibinfo{pages}{4097--4114}.
\newblock
\showISBNx{1063902010}
\showISSN{15737608}
\urldef\tempurl%
\url{https://doi.org/10.1007/s10639-020-10164-w}
\showDOI{\tempurl}


\bibitem[Denny et~al\mbox{.}(2019)]%
        {denny2019research}
\bibfield{author}{\bibinfo{person}{Paul Denny}, \bibinfo{person}{Brett~A.
  Becker}, \bibinfo{person}{Michelle Craig}, \bibinfo{person}{Greg Wilson},
  {and} \bibinfo{person}{Piotr Banaszkiewicz}.}
  \bibinfo{year}{2019}\natexlab{}.
\newblock \showarticletitle{Research This! Questions That Computing Educators
  Most Want Computing Education Researchers to Answer}. In
  \bibinfo{booktitle}{\emph{Proc. of the 2019 ACM Conf. on International
  Computing Education Research}} (Toronto ON, Canada)
  \emph{(\bibinfo{series}{ICER '19})}. \bibinfo{publisher}{ACM},
  \bibinfo{address}{New York, NY, USA}, \bibinfo{pages}{259–267}.
\newblock
\showISBNx{9781450361859}
\urldef\tempurl%
\url{https://doi.org/10.1145/3291279.3339402}
\showDOI{\tempurl}


\bibitem[Dong et~al\mbox{.}(2019)]%
        {dong2019defining}
\bibfield{author}{\bibinfo{person}{Yihuan Dong}, \bibinfo{person}{Samiha
  Marwan}, \bibinfo{person}{Veronica Catete}, \bibinfo{person}{Thomas Price},
  {and} \bibinfo{person}{Tiffany Barnes}.} \bibinfo{year}{2019}\natexlab{}.
\newblock \showarticletitle{Defining Tinkering Behavior in Open-Ended
  Block-Based Programming Assignments}. In \bibinfo{booktitle}{\emph{Proc. of
  the 50th ACM Technical Symposium on Computer Science Education}}
  (Minneapolis, MN, USA) \emph{(\bibinfo{series}{SIGCSE '19})}.
  \bibinfo{publisher}{ACM}, \bibinfo{address}{New York, NY, USA},
  \bibinfo{pages}{1204–1210}.
\newblock
\showISBNx{9781450358903}
\urldef\tempurl%
\url{https://doi.org/10.1145/3287324.3287437}
\showDOI{\tempurl}


\bibitem[Duncan and Bell(2015)]%
        {duncan2015pilot}
\bibfield{author}{\bibinfo{person}{Caitlin Duncan} {and} \bibinfo{person}{Tim
  Bell}.} \bibinfo{year}{2015}\natexlab{}.
\newblock \showarticletitle{A Pilot Computer Science and Programming Course for
  Primary School Students}. In \bibinfo{booktitle}{\emph{Proc. of the Workshop
  in Primary and Secondary Computing Education}} (London, UK)
  \emph{(\bibinfo{series}{WiPSCE '15})}. \bibinfo{publisher}{ACM},
  \bibinfo{address}{New York, NY, USA}, \bibinfo{pages}{39–48}.
\newblock
\showISBNx{9781450337533}
\urldef\tempurl%
\url{https://doi.org/10.1145/2818314.2818328}
\showDOI{\tempurl}


\bibitem[Falkner et~al\mbox{.}(2019)]%
        {falkner2019international}
\bibfield{author}{\bibinfo{person}{Katrina Falkner}, \bibinfo{person}{Sue
  Sentance}, \bibinfo{person}{Rebecca Vivian}, \bibinfo{person}{Sarah
  Barksdale}, \bibinfo{person}{Leonard Busuttil}, \bibinfo{person}{Elizabeth
  Cole}, \bibinfo{person}{Christine Liebe}, \bibinfo{person}{Francesco
  Maiorana}, \bibinfo{person}{Monica~M. McGill}, {and} \bibinfo{person}{Keith
  Quille}.} \bibinfo{year}{2019}\natexlab{}.
\newblock \showarticletitle{An International Comparison of K-12 Computer
  Science Education Intended and Enacted Curricula}. In
  \bibinfo{booktitle}{\emph{Proc. of the 19th Koli Calling Int. Conf. on
  Computing Education Research}} (Koli, Finland) \emph{(\bibinfo{series}{Koli
  Calling '19})}. \bibinfo{publisher}{ACM}, \bibinfo{address}{New York, NY,
  USA}, Article \bibinfo{articleno}{4}, \bibinfo{numpages}{10}~pages.
\newblock
\showISBNx{9781450377157}
\urldef\tempurl%
\url{https://doi.org/10.1145/3364510.3364517}
\showDOI{\tempurl}


\bibitem[(Firm)(2016)]%
        {google2016diversity}
\bibfield{author}{\bibinfo{person}{Google (Firm)~Gallup (Firm)}.}
  \bibinfo{year}{2016}\natexlab{}.
\newblock \bibinfo{title}{Diversity gaps in computer science: exploring the
  underrepresentation of girls, Blacks and Hispanics}.
\newblock
\newblock


\bibitem[Fong et~al\mbox{.}(2021)]%
        {Fong2021}
\bibfield{author}{\bibinfo{person}{Carlton~J. Fong},
  \bibinfo{person}{Kristen~P. Kremer}, \bibinfo{person}{Christie {Hill-Troglin
  Cox}}, {and} \bibinfo{person}{Christie~A. Lawson}.}
  \bibinfo{year}{2021}\natexlab{}.
\newblock \showarticletitle{{Expectancy-value profiles in math and science: A
  person-centered approach to cross-domain motivation with academic and
  STEM-related outcomes}}.
\newblock \bibinfo{journal}{\emph{Contemporary Educational Psychology}}
  \bibinfo{volume}{65}, \bibinfo{number}{March} (\bibinfo{year}{2021}),
  \bibinfo{pages}{101962}.
\newblock
\showISSN{10902384}
\urldef\tempurl%
\url{https://doi.org/10.1016/j.cedpsych.2021.101962}
\showDOI{\tempurl}


\bibitem[Fouad and Santana(2017)]%
        {Fouad2017}
\bibfield{author}{\bibinfo{person}{Nadya~A. Fouad} {and}
  \bibinfo{person}{Mercedes~C. Santana}.} \bibinfo{year}{2017}\natexlab{}.
\newblock \showarticletitle{{SCCT and Underrepresented Populations in STEM
  Fields: Moving the Needle}}.
\newblock \bibinfo{journal}{\emph{Journal of Career Assessment}}
  \bibinfo{volume}{25}, \bibinfo{number}{1} (\bibinfo{year}{2017}),
  \bibinfo{pages}{24--39}.
\newblock
\showISSN{15524590}
\urldef\tempurl%
\url{https://doi.org/10.1177/1069072716658324}
\showDOI{\tempurl}


\bibitem[Frank(1988)]%
        {Frank1988}
\bibfield{author}{\bibinfo{person}{Celeste~P. Frank}.}
  \bibinfo{year}{1988}\natexlab{}.
\newblock \showarticletitle{{the Development of Academic Advising Programs}}.
\newblock \bibinfo{journal}{\emph{NACADA Journal}} \bibinfo{volume}{8},
  \bibinfo{number}{1} (\bibinfo{year}{1988}), \bibinfo{pages}{11--28}.
\newblock
\showISSN{0271-9517}
\urldef\tempurl%
\url{https://doi.org/10.12930/0271-9517-8.1.11}
\showDOI{\tempurl}


\bibitem[Gaddam et~al\mbox{.}(2018)]%
        {gaddam2018culturally}
\bibfield{author}{\bibinfo{person}{Anuroop Gaddam}, \bibinfo{person}{Karsten
  Lundqvist}, {and} \bibinfo{person}{Hiria McRae}.}
  \bibinfo{year}{2018}\natexlab{}.
\newblock \showarticletitle{Culturally Relevant Approach to Encourage School
  Children Learn Computer Science Concepts}. In \bibinfo{booktitle}{\emph{2018
  Int. Conf. on Learning and Teaching in Computing and Engineering (LaTICE)}}.
  \bibinfo{publisher}{IEEE Computer Society}, \bibinfo{address}{USA},
  \bibinfo{pages}{6--10}.
\newblock
\urldef\tempurl%
\url{https://doi.org/10.1109/LaTICE.2018.00009}
\showDOI{\tempurl}


\bibitem[G{\"{o}}k{\c{c}}earslan et~al\mbox{.}(2017)]%
        {Gokcearslan2017}
\bibfield{author}{\bibinfo{person}{Şahin G{\"{o}}k{\c{c}}earslan},
  \bibinfo{person}{Mustafa~Serkan G{\"{u}}nbatar}, {and}
  \bibinfo{person}{Volkan Kukul}.} \bibinfo{year}{2017}\natexlab{}.
\newblock \showarticletitle{{Computer Programming Self-Efficacy Scale (CPSES)
  for Secondary School Students: Development, Validation and Reliability}}.
\newblock \bibinfo{journal}{\emph{Eğitim Teknolojisi Kuram ve Uygulama}}
  \bibinfo{volume}{7}, \bibinfo{number}{1} (\bibinfo{year}{2017}),
  \bibinfo{pages}{158--158}.
\newblock
\showISBNx{2147-1908}
\showISSN{2147-1908}
\urldef\tempurl%
\url{https://doi.org/10.17943/etku.288493}
\showDOI{\tempurl}


\bibitem[Guzdial et~al\mbox{.}(2012)]%
        {Guzdial2012}
\bibfield{author}{\bibinfo{person}{Mark Guzdial}, \bibinfo{person}{Barbara~J.
  Ericson}, \bibinfo{person}{Tom McKlin}, {and} \bibinfo{person}{Shelly
  Engelman}.} \bibinfo{year}{2012}\natexlab{}.
\newblock \showarticletitle{A Statewide Survey on Computing Education Pathways
  and Influences: Factors in Broadening Participation in Computing}. In
  \bibinfo{booktitle}{\emph{Proc. of the Ninth Annual Int. Conf. on
  International Computing Education Research}} (Auckland, New Zealand)
  \emph{(\bibinfo{series}{ICER '12})}. \bibinfo{publisher}{ACM},
  \bibinfo{address}{New York, NY, USA}, \bibinfo{pages}{143–150}.
\newblock
\showISBNx{9781450316040}
\urldef\tempurl%
\url{https://doi.org/10.1145/2361276.2361304}
\showDOI{\tempurl}


\bibitem[Hanson(2004)]%
        {Hanson2004}
\bibfield{author}{\bibinfo{person}{S. Hanson}.}
  \bibinfo{year}{2004}\natexlab{}.
\newblock \showarticletitle{{African American Women in Science : Experiences
  from High School through the Post- Secondary Years and Beyond Author(s):
  Sandra L . Hanson Published by : The Johns Hopkins University Press Stable
  URL : http://www.jstor.org/stable/4317036}}.
\newblock \bibinfo{journal}{\emph{NWSA Journal}} \bibinfo{volume}{16},
  \bibinfo{number}{1} (\bibinfo{year}{2004}), \bibinfo{pages}{96--115}.
\newblock


\bibitem[Kesselbacher and Bollin(2019)]%
        {Kesselbacher2019}
\bibfield{author}{\bibinfo{person}{Max Kesselbacher} {and}
  \bibinfo{person}{Andreas Bollin}.} \bibinfo{year}{2019}\natexlab{}.
\newblock \showarticletitle{{Quantifying patterns and programming strategies in
  block-based programming environments}}, In \bibinfo{booktitle}{2019 IEEE/ACM
  41st Int. Conf. on Software Engineering: Companion Proceedings
  (ICSE-Companion)}.
\newblock \bibinfo{journal}{\emph{Proceedings - 2019 IEEE/ACM 41st Int. Conf.
  on Software Engineering: Companion, ICSE-Companion 2019}},
  \bibinfo{pages}{254--255}.
\newblock
\showISBNx{9781728117645}
\urldef\tempurl%
\url{https://doi.org/10.1109/ICSE-Companion.2019.00101}
\showDOI{\tempurl}


\bibitem[Klassner and Anderson(2003)]%
        {klassner2003lego}
\bibfield{author}{\bibinfo{person}{F. Klassner} {and} \bibinfo{person}{S.D.
  Anderson}.} \bibinfo{year}{2003}\natexlab{}.
\newblock \showarticletitle{LEGO MindStorms: not just for K-12 anymore}.
\newblock \bibinfo{journal}{\emph{IEEE Robotics Automation Magazine}}
  \bibinfo{volume}{10}, \bibinfo{number}{2} (\bibinfo{year}{2003}),
  \bibinfo{pages}{12--18}.
\newblock
\urldef\tempurl%
\url{https://doi.org/10.1109/MRA.2003.1213611}
\showDOI{\tempurl}


\bibitem[Korkmaz et~al\mbox{.}(2017)]%
        {Korkmaz2017}
\bibfield{author}{\bibinfo{person}{{\"{O}}zgen Korkmaz}, \bibinfo{person}{Recep
  {\c{C}}akir}, {and} \bibinfo{person}{M.~Yaşar {\"{O}}zden}.}
  \bibinfo{year}{2017}\natexlab{}.
\newblock \showarticletitle{{A validity and reliability study of the
  computational thinking scales (CTS)}}.
\newblock \bibinfo{journal}{\emph{Computers in Human Behavior}}
  \bibinfo{volume}{72} (\bibinfo{year}{2017}), \bibinfo{pages}{558--569}.
\newblock
\showISSN{07475632}
\urldef\tempurl%
\url{https://doi.org/10.1016/j.chb.2017.01.005}
\showDOI{\tempurl}


\bibitem[Lent et~al\mbox{.}(1994)]%
        {Lent1994}
\bibfield{author}{\bibinfo{person}{Robert Lent}, \bibinfo{person}{Steven
  Brown}, {and} \bibinfo{person}{Gail Hackett}.}
  \bibinfo{year}{1994}\natexlab{}.
\newblock \showarticletitle{Toward a Unifying Social Cognitive Theory of Career
  and Academic Interest, Choice, and Performance}.
\newblock \bibinfo{journal}{\emph{J. of Vocational Behavior}}
  \bibinfo{volume}{45} (\bibinfo{date}{08} \bibinfo{year}{1994}),
  \bibinfo{pages}{79--122}.
\newblock
\urldef\tempurl%
\url{https://doi.org/10.1006/jvbe.1994.1027}
\showDOI{\tempurl}


\bibitem[Loksa et~al\mbox{.}(2022)]%
        {loksa2022metacognition}
\bibfield{author}{\bibinfo{person}{Dastyni Loksa}, \bibinfo{person}{Lauren
  Margulieux}, \bibinfo{person}{Brett~A. Becker}, \bibinfo{person}{Michelle
  Craig}, \bibinfo{person}{Paul Denny}, \bibinfo{person}{Raymond Pettit}, {and}
  \bibinfo{person}{James Prather}.} \bibinfo{year}{2022}\natexlab{}.
\newblock \showarticletitle{Metacognition and Self-Regulation in Programming
  Education: Theories and Exemplars of Use}.
\newblock \bibinfo{journal}{\emph{ACM Trans. Comput. Educ.}}
  \bibinfo{volume}{22}, \bibinfo{number}{4}, Article \bibinfo{articleno}{39}
  (\bibinfo{date}{sep} \bibinfo{year}{2022}), \bibinfo{numpages}{31}~pages.
\newblock
\urldef\tempurl%
\url{https://doi.org/10.1145/3487050}
\showDOI{\tempurl}


\bibitem[Lunn et~al\mbox{.}(2021)]%
        {lunn2021exploration}
\bibfield{author}{\bibinfo{person}{Stephanie Lunn}, \bibinfo{person}{Leila
  Zahedi}, \bibinfo{person}{Monique Ross}, {and} \bibinfo{person}{Matthew
  Ohland}.} \bibinfo{year}{2021}\natexlab{}.
\newblock \showarticletitle{Exploration of Intersectionality and Computer
  Science Demographics: Understanding the Historical Context of Shifts in
  Participation}.
\newblock \bibinfo{journal}{\emph{ACM Trans. Comput. Educ.}}
  \bibinfo{volume}{21}, \bibinfo{number}{2}, Article \bibinfo{articleno}{10}
  (\bibinfo{date}{mar} \bibinfo{year}{2021}), \bibinfo{numpages}{30}~pages.
\newblock
\urldef\tempurl%
\url{https://doi.org/10.1145/3445985}
\showDOI{\tempurl}


\bibitem[Margolis et~al\mbox{.}(2012)]%
        {margolis2012beyond}
\bibfield{author}{\bibinfo{person}{Jane Margolis}, \bibinfo{person}{Jean~J.
  Ryoo}, \bibinfo{person}{Cueponcaxochitl D.~M. Sandoval},
  \bibinfo{person}{Clifford Lee}, \bibinfo{person}{Joanna Goode}, {and}
  \bibinfo{person}{Gail Chapman}.} \bibinfo{year}{2012}\natexlab{}.
\newblock \showarticletitle{Beyond Access: Broadening Participation in High
  School Computer Science}.
\newblock \bibinfo{journal}{\emph{ACM Inroads}} \bibinfo{volume}{3},
  \bibinfo{number}{4} (\bibinfo{date}{dec} \bibinfo{year}{2012}),
  \bibinfo{pages}{72–78}.
\newblock
\showISSN{2153-2184}
\urldef\tempurl%
\url{https://doi.org/10.1145/2381083.2381102}
\showDOI{\tempurl}


\bibitem[Martin et~al\mbox{.}(2021)]%
        {Martin2021}
\bibfield{author}{\bibinfo{person}{Andrew~J. Martin}, \bibinfo{person}{Roger
  Kennett}, \bibinfo{person}{Joel Pearson}, \bibinfo{person}{Marianne Mansour},
  \bibinfo{person}{Brad Papworth}, {and} \bibinfo{person}{Lars~Erik Malmberg}.}
  \bibinfo{year}{2021}\natexlab{}.
\newblock \showarticletitle{{Challenge and threat appraisals in high school
  science: investigating the roles of psychological and physiological
  factors}}.
\newblock \bibinfo{journal}{\emph{Educational Psychology}}
  \bibinfo{volume}{41}, \bibinfo{number}{5} (\bibinfo{year}{2021}),
  \bibinfo{pages}{618--639}.
\newblock
\showISSN{14695820}
\urldef\tempurl%
\url{https://doi.org/10.1080/01443410.2021.1887456}
\showDOI{\tempurl}


\bibitem[McAllister et~al\mbox{.}(2022)]%
        {McAllister2022}
\bibfield{author}{\bibinfo{person}{Tara~G McAllister}, \bibinfo{person}{Sereana
  Naepi}, \bibinfo{person}{Elizabeth Wilson}, \bibinfo{person}{Daniel Hikuroa},
  {and} \bibinfo{person}{Leilani~A Walker}.} \bibinfo{year}{2022}\natexlab{}.
\newblock \showarticletitle{Under-represented and overlooked: M{\=a}ori and
  Pasifika scientists in Aotearoa New Zealand’s universities and
  crown-research institutes}.
\newblock \bibinfo{journal}{\emph{Journal of the Royal Society of New Zealand}}
  \bibinfo{volume}{52}, \bibinfo{number}{1} (\bibinfo{year}{2022}),
  \bibinfo{pages}{38--53}.
\newblock


\bibitem[McBroom et~al\mbox{.}(2020)]%
        {mcbroom2020understanding}
\bibfield{author}{\bibinfo{person}{Jessica McBroom}, \bibinfo{person}{Irena
  Koprinska}, {and} \bibinfo{person}{Kalina Yacef}.}
  \bibinfo{year}{2020}\natexlab{}.
\newblock \showarticletitle{Understanding Gender Differences to Improve Equity
  in Computer Programming Education}. In \bibinfo{booktitle}{\emph{Proc. of the
  Twenty-Second Australasian Computing Education Conf.}} (Melbourne, Australia)
  \emph{(\bibinfo{series}{ACE'20})}. \bibinfo{publisher}{ACM},
  \bibinfo{address}{NY, USA}, \bibinfo{pages}{185–194}.
\newblock
\showISBNx{9781450376860}
\urldef\tempurl%
\url{https://doi.org/10.1145/3373165.3373186}
\showDOI{\tempurl}


\bibitem[Mellis et~al\mbox{.}(2007)]%
        {mellis2007arduino}
\bibfield{author}{\bibinfo{person}{David Mellis}, \bibinfo{person}{Massimo
  Banzi}, \bibinfo{person}{David Cuartielles}, {and} \bibinfo{person}{Tom
  Igoe}.} \bibinfo{year}{2007}\natexlab{}.
\newblock \showarticletitle{Arduino: An open electronic prototyping platform}.
  In \bibinfo{booktitle}{\emph{Proc. Chi}}, Vol.~\bibinfo{volume}{2007}.
  \bibinfo{pages}{1--11}.
\newblock


\bibitem[Newhall et~al\mbox{.}(2014)]%
        {newhall2014support}
\bibfield{author}{\bibinfo{person}{Tia Newhall}, \bibinfo{person}{Lisa Meeden},
  \bibinfo{person}{Andrew Danner}, \bibinfo{person}{Ameet Soni},
  \bibinfo{person}{Frances Ruiz}, {and} \bibinfo{person}{Richard Wicentowski}.}
  \bibinfo{year}{2014}\natexlab{}.
\newblock \showarticletitle{A Support Program for Introductory CS Courses That
  Improves Student Performance and Retains Students from Underrepresented
  Groups}. In \bibinfo{booktitle}{\emph{Proc. of the 45th ACM Technical
  Symposium on Computer Science Education}} (Atlanta, Georgia, USA)
  \emph{(\bibinfo{series}{SIGCSE '14})}. \bibinfo{publisher}{ACM},
  \bibinfo{address}{New York, NY, USA}, \bibinfo{pages}{433–438}.
\newblock
\showISBNx{9781450326056}
\urldef\tempurl%
\url{https://doi.org/10.1145/2538862.2538923}
\showDOI{\tempurl}


\bibitem[Percival et~al\mbox{.}(2022)]%
        {Percival2022}
\bibfield{author}{\bibinfo{person}{Nathan Percival}, \bibinfo{person}{Pranathi
  Rayavaram}, \bibinfo{person}{Sashank Narain}, {and}
  \bibinfo{person}{Claire~Seungeun Lee}.} \bibinfo{year}{2022}\natexlab{}.
\newblock \showarticletitle{{CryptoScratch: Developing and evaluating a
  block-based programming tool for teaching K-12 cryptography education using
  Scratch}}.
\newblock \bibinfo{journal}{\emph{IEEE Global Engineering Education Conf.,
  EDUCON}}  \bibinfo{volume}{2022-March} (\bibinfo{year}{2022}),
  \bibinfo{pages}{1004--1013}.
\newblock
\showISBNx{9781665444347}
\showISSN{21659567}
\urldef\tempurl%
\url{https://doi.org/10.1109/EDUCON52537.2022.9766637}
\showDOI{\tempurl}


\bibitem[Rachmatullah et~al\mbox{.}(2020)]%
        {Rachmatullah2020}
\bibfield{author}{\bibinfo{person}{Arif Rachmatullah}, \bibinfo{person}{Eric
  Wiebe}, \bibinfo{person}{Danielle Boulden}, \bibinfo{person}{Bradford Mott},
  \bibinfo{person}{Kristy Boyer}, {and} \bibinfo{person}{James Lester}.}
  \bibinfo{year}{2020}\natexlab{}.
\newblock \showarticletitle{{Development and validation of the Computer Science
  Attitudes Scale for middle school students (MG-CS attitudes)}}.
\newblock \bibinfo{journal}{\emph{Computers in Human Behavior Reports}}
  \bibinfo{volume}{2}, \bibinfo{number}{May} (\bibinfo{year}{2020}).
\newblock
\showISSN{24519588}
\urldef\tempurl%
\url{https://doi.org/10.1016/j.chbr.2020.100018}
\showDOI{\tempurl}


\bibitem[Schlegel et~al\mbox{.}(2019)]%
        {Schlegel2019}
\bibfield{author}{\bibinfo{person}{Rebecca~J. Schlegel},
  \bibinfo{person}{Sharon~L. Chu}, \bibinfo{person}{Kaiyuan Chen},
  \bibinfo{person}{Elizabeth Deuermeyer}, \bibinfo{person}{Andrew~G. Christy},
  {and} \bibinfo{person}{Francis Quek}.} \bibinfo{year}{2019}\natexlab{}.
\newblock \showarticletitle{{Making in the classroom: Longitudinal evidence of
  increases in self-efficacy and STEM possible selves over time}}.
\newblock \bibinfo{journal}{\emph{Computers and Education}}
  \bibinfo{volume}{142}, \bibinfo{number}{July} (\bibinfo{year}{2019}).
\newblock
\showISSN{03601315}
\urldef\tempurl%
\url{https://doi.org/10.1016/j.compedu.2019.103637}
\showDOI{\tempurl}


\bibitem[Sentance et~al\mbox{.}(2017)]%
        {sentance2017creating}
\bibfield{author}{\bibinfo{person}{Sue Sentance}, \bibinfo{person}{Jane Waite},
  \bibinfo{person}{Steve Hodges}, \bibinfo{person}{Emily MacLeod}, {and}
  \bibinfo{person}{Lucy Yeomans}.} \bibinfo{year}{2017}\natexlab{}.
\newblock \showarticletitle{"Creating Cool Stuff": Pupils' Experience of the
  BBC Micro:Bit}. In \bibinfo{booktitle}{\emph{Proc. of the 2017 ACM SIGCSE
  Tech. Symp. on CS Education}} (Seattle, Washington, USA)
  \emph{(\bibinfo{series}{SIGCSE '17})}. \bibinfo{publisher}{ACM},
  \bibinfo{address}{NY, USA}, \bibinfo{pages}{531–536}.
\newblock
\showISBNx{9781450346986}
\urldef\tempurl%
\url{https://doi.org/10.1145/3017680.3017749}
\showDOI{\tempurl}


\bibitem[Suriyaarachchi et~al\mbox{.}(2022b)]%
        {suriyaarachchi2022primary}
\bibfield{author}{\bibinfo{person}{Hussel Suriyaarachchi},
  \bibinfo{person}{Paul Denny}, \bibinfo{person}{Juan Pablo~Forero Cortes},
  \bibinfo{person}{Chamod Weerasinghe}, {and} \bibinfo{person}{Suranga
  Nanayakkara}.} \bibinfo{year}{2022}\natexlab{b}.
\newblock \showarticletitle{Primary School Students Programming with Real-Time
  Environmental Sensor Data}. In \bibinfo{booktitle}{\emph{Proc. of the 24th
  Australasian Computing Education Conference}} (Virtual Event, Australia)
  \emph{(\bibinfo{series}{ACE '22})}. \bibinfo{publisher}{ACM},
  \bibinfo{address}{New York, NY, USA}, \bibinfo{pages}{85–94}.
\newblock
\showISBNx{9781450396431}
\urldef\tempurl%
\url{https://doi.org/10.1145/3511861.3511871}
\showDOI{\tempurl}


\bibitem[Suriyaarachchi et~al\mbox{.}(2022a)]%
        {Suriyaarachchi2022}
\bibfield{author}{\bibinfo{person}{Hussel Suriyaarachchi},
  \bibinfo{person}{Paul Denny}, {and} \bibinfo{person}{Suranga Nanayakkara}.}
  \bibinfo{year}{2022}\natexlab{a}.
\newblock \showarticletitle{Scratch and Sense: Using Real-Time Sensor Data to
  Motivate Students Learning Scratch}. In \bibinfo{booktitle}{\emph{Proc. of
  the 53rd ACM Technical Symposium on Computer Science Education - Volume 1}}
  (Providence, RI, USA) \emph{(\bibinfo{series}{SIGCSE 2022})}.
  \bibinfo{publisher}{ACM}, \bibinfo{address}{New York, NY, USA},
  \bibinfo{pages}{983–989}.
\newblock
\showISBNx{9781450390705}
\urldef\tempurl%
\url{https://doi.org/10.1145/3478431.3499316}
\showDOI{\tempurl}


\bibitem[Tsai et~al\mbox{.}(2019)]%
        {Tsai2019}
\bibfield{author}{\bibinfo{person}{Meng~Jung Tsai}, \bibinfo{person}{Ching~Yeh
  Wang}, {and} \bibinfo{person}{Po~Fen Hsu}.} \bibinfo{year}{2019}\natexlab{}.
\newblock \showarticletitle{{Developing the Computer Programming Self-Efficacy
  Scale for Computer Literacy Education}}.
\newblock \bibinfo{journal}{\emph{Journal of Educational Computing Research}}
  \bibinfo{volume}{56}, \bibinfo{number}{8} (\bibinfo{year}{2019}),
  \bibinfo{pages}{1345--1360}.
\newblock
\showISSN{15414140}
\urldef\tempurl%
\url{https://doi.org/10.1177/0735633117746747}
\showDOI{\tempurl}


\bibitem[Vieira(2016)]%
        {vieira2016t}
\bibfield{author}{\bibinfo{person}{Pedro~Cosme Vieira}.}
  \bibinfo{year}{2016}\natexlab{}.
\newblock \showarticletitle{T-test with Likert scale variables}.
\newblock \bibinfo{journal}{\emph{Available at SSRN 2770035}}
  (\bibinfo{year}{2016}).
\newblock


\bibitem[Weintrop et~al\mbox{.}(2018)]%
        {Weintrop2018}
\bibfield{author}{\bibinfo{person}{David Weintrop}, \bibinfo{person}{Heather
  Killen}, {and} \bibinfo{person}{Baker Franke}.}
  \bibinfo{year}{2018}\natexlab{}.
\newblock \showarticletitle{{Blocks or text? How programming language modality
  makes a difference in assessing underrepresented populations}}.
\newblock \bibinfo{journal}{\emph{Proc. of Int. Conf. of the Learning Sciences,
  ICLS}} \bibinfo{volume}{1}, \bibinfo{number}{2018-June}
  (\bibinfo{year}{2018}), \bibinfo{pages}{328--335}.
\newblock
\showISSN{18149316}


\bibitem[Wiebe et~al\mbox{.}(2003)]%
        {Wiebe2003}
\bibfield{author}{\bibinfo{person}{Eric Wiebe}, \bibinfo{person}{Laurie~Ann
  Williams}, \bibinfo{person}{Kai Yang}, {and} \bibinfo{person}{Carol~S
  Miller}.} \bibinfo{year}{2003}\natexlab{}.
\newblock \bibinfo{booktitle}{\emph{Computer Science Attitude Survey}}.
\newblock \bibinfo{type}{{T}echnical {R}eport}. \bibinfo{institution}{North
  Carolina State University. Dept. of Computer Science}.
\newblock


\bibitem[Yuen et~al\mbox{.}(2016)]%
        {yuen2016culturally}
\bibfield{author}{\bibinfo{person}{Timothy Yuen}, \bibinfo{person}{Maria
  Arreguin-Anderson}, \bibinfo{person}{Guadalupe Carmona}, {and}
  \bibinfo{person}{Matthew Gibson}.} \bibinfo{year}{2016}\natexlab{}.
\newblock \showarticletitle{A Culturally Relevant Pedagogical Approach to
  Computer Science Education to Increase Participation of Underrepresented
  Populations}. In \bibinfo{booktitle}{\emph{2016 Int. Conf. on Learning and
  Teaching in Computing and Engineering (LaTICE)}}. \bibinfo{pages}{147--153}.
\newblock
\urldef\tempurl%
\url{https://doi.org/10.1109/LaTiCE.2016.44}
\showDOI{\tempurl}


\end{thebibliography}

\end{document}